 \documentclass[final,5p,times,twocolumn,authoryear]{elsarticle}

\usepackage{amssymb}
\usepackage{lipsum}
\usepackage[utf8]{inputenc}
\usepackage{amsmath}
\usepackage{hyperref}
\usepackage{graphicx}
\usepackage{color}
\usepackage{ulem}
\usepackage{booktabs}
\journal{High Energy Astrophysics}

\begin{document}

\begin{frontmatter}



\title{Tidal Properties of White Dwarfs Admixed with Fermionic Dark Matter}


\author[first,b,c]{G. A. Carvalho}
\affiliation[first]{organization={Departamento de Física, Universidade Tecnológica Federal do Paraná},%Department and Organization
            addressline={Avenida Brasil 4232, Parque Independência}, 
            city={Medianeira},
            postcode={85722-332}, 
            state={PR},
            country={Brazil}}

\affiliation[b]{organization={Programa de P\'os-Gradua\c{c}\~ao em F\'isica e Astronomia, Universidade Tecnol\'ogica Federal do Paran\'a},
            addressline={ Av. Sete de Setembro 3165, Rebouças}, 
            city={Curitiba},
            postcode={80230-901}, 
            state={PR},
            country={Brasil}}

\affiliation[c]{organization={Núcleo de Astrofísica e Cosmologia (Cosmo-Ufes) \& Departamento de Física},
            addressline={Avenida Fernando Ferrari 514, Goiabeiras}, 
            city={Vitória},
            postcode={29075-910}, 
            state={ES},
            country={Brasil}}

\author[label2,label3]{J. D. V. Arba\~nil}
\affiliation[label2]{organization={Departamento de Ciencias, Universidad Privada del Norte, Avenida el Sol 461 San Juan de Lurigancho, 15434 Lima,  Peru}}
\affiliation[label3]{organization={Facultad de Ciencias Físicas, Universidad Nacional Mayor de San Marcos, Avenida Venezuela s/n Cercado de Lima, 15081 Lima,  Peru}}
\author[c]{I. A. Lacerda}
\author[c,e]{Jaziel G. Coelho}
\address[e]{Divisão de Astrofisica, Instituto Nacional de Pesquisas Espaciais, Avenida dos Astronautas 1758, 12227-010, São José dos Campos, SP, Brazil}

\begin{abstract}
This study investigates the impact of fermionic dark matter (DM) on the structural and tidal properties of white dwarfs (WDs). Using a two-fluid general relativistic framework, we model stars with fixed DM mass fractions (${\rm MF}=M_{\rm DM}/M_{\rm NM} $), ranging from 0\% to 8\%, with zero representing the no DM scenario. Our results indicate that DM accumulation leads to a significant compactification of the star, reflected in increased surface gravity and compactness. Furthermore, we calculate the second-order tidal Love number $k_2$ and the dimensionless tidal deformability $\Lambda$. For some total mass range, we find that the presence of DM significantly reduces $\Lambda$, suggesting that DM-admixed WDs (DMWDs) are less susceptible to tidal deformations. These structural shifts alter the gravitational wave signatures in binary systems, potentially providing a new window for DM detection with future instruments like LISA.
\end{abstract}




\end{frontmatter}

%\tableofcontents



\section{Introduction}
WDs are among the best laboratories for studying matter under extreme conditions because their equation of state is well constrained and their equilibrium structure is accurately described within general relativity in contrast with the description of the neutron star equation of state. This makes them suitable for testing compact-star physics, including the effects of temperature, magnetization, composition, crystallization, and nonstandard matter components\textcolor{blue}{; see, e. g.,}  \citep{Franzon2015-lx,Chatterjee2017-hh,Carvalho2018-wc,Otoniel2019-vm,Nunes2021-lb,Camisassa2022-xt,Althaus2022-qd,Carvalho2025,Nunes2025,Sahoo2025-ac,Kalita2026-nq,Arseneau2026-zr,Lobato2026-ba,Ridha_Fathima2026-po,Malheiro2026-xu,Nunes2026-pg,Glanz2027-is}. Recent work has shown that WDs can also serve as probes of DM through the gravitational impact of an admixed dark component \citep{Carvalho2025,Nunes2026-pg,Sahoo2026-ih}.

In a previous study, \cite{Carvalho2025} investigated WDs admixed with light fermionic DM within a relativistic two-fluid framework. In this approach, the ordinary and dark-matter components are treated as distinct fluids coupled only through gravity, allowing their individual density and pressure profiles to be determined self-consistently. For fermionic dark-matter particle masses in the range 0.1-10 GeV, the authors found that even a relatively small dark-matter contribution can substantially modify the equilibrium sequence of WDs. In particular, the presence of the dark component may lead to smaller stellar radii and larger compactnesses for a given total mass, thereby shifting the standard mass–radius relation toward more compact configurations. Similar findings are discussed by \cite{Sahoo2026-ih}, where increasing the DM fraction inside WDs makes the stars more compact, with smaller masses and radii.

These structural changes are accompanied by modifications to the stability properties of the stars, which depend on the dark-matter fraction, particle mass, and central-density distribution. The resulting configurations may also exhibit distinctive changes in their surface gravity and gravitational redshift, providing potential observational signatures of DM inside WDs \citep{Carvalho2025}. More generally, such effects suggest that precise measurements of white-dwarf masses, radii, and spectroscopic properties could be used to constrain the allowed amount and particle properties of fermionic DM, although degeneracies with temperature, rotation, magnetic fields, and uncertainties in the stellar equation of state must also be taken into account.

In a complementary direction, future space-based gravitational-wave detectors will provide a unique opportunity to probe additional properties of WDs through the imprints of tidal interactions on the gravitational-wave signals emitted by close or merging white-dwarf binaries. In particular, the tidal response of these systems may offer valuable information about the internal composition and thermodynamic state of WDs, including the possible presence of crystallized cores \cite{Perot2022-uh,Tang2023-jw}. By measuring tidal effects in the inspiral waveform, it may therefore be possible to place constraints on the equation of state and internal structure of these compact objects. 

In this context, \cite{Nunes2025} investigated the tidal response of WDs and computed their Love numbers, tidal deformabilities, and gravitational-wave properties, considering hot WDs in white-dwarf binaries. Their results demonstrate that finite-temperature effects can significantly influence the tidal interaction and, consequently, the gravitational-wave signal emitted by white-dwarf binaries. This highlights the importance of incorporating thermal profiles when modeling such systems and interpreting future gravitational-wave observations.

These two lines of research, namely, DM and tidal interactions, motivate the present work. Here we combine the tidal-deformability formalism with the two-fluid fermionic dark-matter model in order to study WDs with a fixed dark-matter mass fraction. Instead of parameterizing the dark component through a central Fermi momentum, we impose a fixed mass ratio $M_{\rm DM}/M_{\rm WD}$. This allows a direct assessment of how the dark-matter content modifies the global structure and tidal properties of WDs.

The paper is organized as follows. Section \ref{sec2} presents the model and the structure equations. Section \ref{sec3} discusses the mass-radius relation, Love number, deformability, surface gravity, and binary diagrams. Section \ref{sec4} summarizes the main conclusions. Throughout this work, we adopt geometrized units, setting $G=1=c$, to simplify the equations and facilitate the numerical calculations.

\section{Equation of state, static equilibrium structure equations, and tidal deformability equations}
\label{sec2}
\subsection{Equation of state}

To investigate the effects of DM on the equilibrium structure and tidal deformability of WDs, we describe the ordinary-matter component using the relativistic degenerate-electron equation of state, including the ionic rest-mass contribution, as used by \cite{Carvalho2025}. This equation of state corresponds to the standard Chandrasekhar model, where the pressure is provided by the degenerate electron gas, while the energy density includes contributions from both the ions and electrons and is dominated by the ionic rest-mass. This is essential for dense white-dwarf matter, where the ionic contribution dominates the mass density even though the degenerate electron gas supports the pressure.

The dark component is modeled as a zero-temperature fermionic gas with its pressure and energy density given by the standard expressions for a relativistic Fermi-gas. In the present work, this component is considered to be non-self-interacting and couples to the normal fluid only through gravity, following the two-fluid formalism as used by \cite{Carvalho2025}.

To investigate the impact of DM on the structural properties of compact stars, the dark-matter mass fraction in each stellar configuration is fixed through the total gravitational mass partition,
\begin{equation}\label{massfrac}
\frac{M_{\rm DM}}{M_{\rm NM}} = {\rm MF},
\end{equation}
with MF ranging from $0.0$ to $0.08$; representing $0$ and $8\%$ of DM inside a white dwarf, respectively. This prescription defines a sequence of equilibrium configurations from the dark-matter-free case to increasingly dark-matter-rich WDs.

\subsection{Two-fluid structure equations}
The spacetime describing the interior of the static, spherically symmetric white dwarf composed of two gravitationally coupled fluids (ordinary white-dwarf matter and fermionic DM) is given by
\begin{equation}
ds^2 = -e^{\nu(r)}dt^2 + e^{\lambda(r)}dr^2 + r^2 \left(d\theta^2+\sin^2\theta d\phi^2\right),
\end{equation}
with
\begin{equation}
e^{-\lambda(r)} = 1 - \frac{2m(r)}{r}.
\end{equation}
The function $m(r)$ represents the mass inside the sphere of radius $r$. The equilibrium solutions  are found by solving the set of two fluid stellar equilibrium equations, known also as Tolman-Oppenheimer-Volkoff equations \citep{2015PhRvD..92l3002T,2016PhRvD..93h3009M}, which are given by,  
\begin{eqnarray}
&&\frac{dp_{i}}{dr} = -\frac{M \epsilon_{i}}{r^2} \left(1 + \frac{p_{i}}{\epsilon_{i}}\right) \left(1 + \frac{4\pi r^3 P}{M}\right) \left(1 - \frac{2M}{r}\right)^{-1},\label{dpdr}\\
&&\frac{dm_i}{dr}=4\pi \epsilon_i r^2,\label{dmdr}
\end{eqnarray}
where $i = \{\rm NM, DM\}$, $P = p_{\rm NM} + p_{\rm DM}$, and $M = m_{\rm NM} + m_{\rm DM}$. The NM represents the normal white dwarf matter and it is described by a relativistic degenerate electron gas (Chandrasekhar model) including ion rest mass, while DM represents the DM contribution and it is modeled as a cold fermionic gas. 

The integration of the set of differential equations, Eqs. \eqref{dpdr} and \eqref{dmdr}, starts at the center of the star $r=0$ where $\epsilon_{\rm NM}(0)=\epsilon_c, p_{\rm NM}(0)=p_c, M(0)=0$ ($m_{\rm NM}(0)=m_{\rm DM}(0)=0$) and ends at the star's surface $r=R$ where both pressures vanish $p_i(R)=0$. The central values for DM energy density and pressure are adapted such that the mass fraction in Eq. \eqref{massfrac} is fixed. Each matter component can reach zero pressure at a different radius, giving rise to distinct NM and DM radii. The stellar radius is then defined as the larger of the two.

\subsection{Love number and tidal deformability}

The tidal response of the star is determined by solving the first-order Riccati differential equation for the auxiliary function $y(r)$, which can be written as \citep{2010PhRvD..82b4016P}
\begin{equation}\label{riccati_de}
    \frac{dy}{dr}= -\frac{y^2}{r}-\frac{y}{r}F-Qr,
\end{equation}
where the functions $F$ and $Q$ are given by
\begin{align}
    F&= [1+4\pi r^2(P-\epsilon)]{\rm e}^{\lambda},\\
    Q&= 4\pi{\rm e}^{\lambda} \left(5\epsilon + 9P + \frac{P+\epsilon}{c_s^2}\right) - \frac{6\rm e^\lambda}{r^2}-\frac{4\rm e^{2\lambda}}{r^4}(M+4\pi r^3P)^2,
\end{align}
with $\epsilon=\epsilon_{\rm NM}+\epsilon_{\rm DM}$. The function $y(r)$ is essential for determining the Love number $k_2$. To this end, the differential equation (\ref{riccati_de}) is integrated simultaneously with the stellar equilibrium equations from the center $r=0$, where $y(r=0)=2$, to the surface of the star $r=R$, whose value of $y(r=R)$ at this point is denoted by $y_R$. Thus, according to \cite{2010PhRvD..82b4016P}, for small compactness $C=M/R<0.1$, the Love number can be calculated approximately as
\begin{equation}
k_2= \frac{1}{2} \frac{(2-y_R)}{(3+y_R)}.
\end{equation}
The compactness of white dwarfs are typically in the order of $10^{-4}-10^{-3}$. By employing the Love number parameter $k_2$, the dimensionless tidal deformability is obtained by the relation:
\begin{equation}
\Lambda = \frac{2}{3}k_2 C^{-5},
\end{equation}
see, for instance, Refs. \cite{2008ApJ...677.1216H,2009PhRvD..80h4035D,2010PhRvD..81l3016H}. In this way, the Riccati differential equation, the Love number equation, and the dimensionless tidal deformability equation allow for investigating the effects of DM on the white dwarf tidal deformability.

\section{Results and discussion}
\label{sec3}

\begin{figure}
    \centering
    \includegraphics[width=1.0\linewidth]{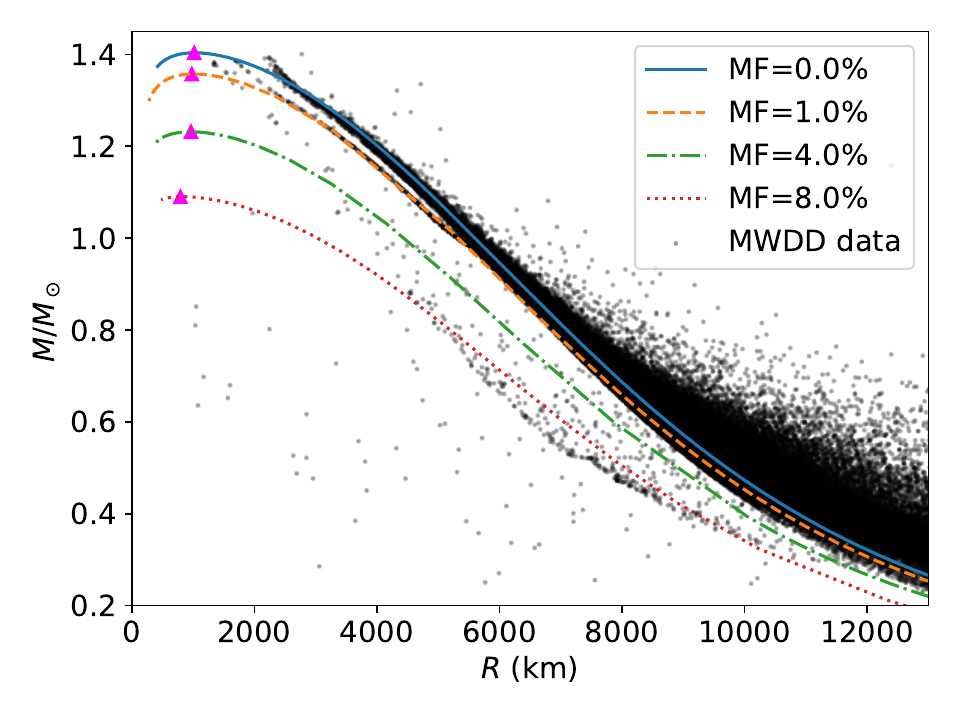}
    \caption{Total mass (in units of solar masses, $M_\odot$) as a function of the total radius for different quantities of DM within WDs. The full pink triangles denote the maximum mass points. Black dots represent observational data taken from the Montreal White Dwarf Database (MWDD) \citep{MWDD}.}
    \label{massxradius}
\end{figure}

The mass as a function of the radius is plotted in Fig. \ref{massxradius} for different DM quantities.  Note that in all curves of the figure, the mass increases as the total radius decreases, until it attains the maximum mass points, marked with full pink triangles. The effects of DM on the equilibrium configurations can be observed in the mass-radius sequence.
The zero-dark-matter case reproduces the standard white-dwarf scenario, with larger radii for low masses and a steep contraction as the mass approaches the Chandrasekhar limit. As the dark-matter fraction increases from ${\rm MF}=1\%$ to $8\%$, the entire sequence shifts toward smaller radii at fixed mass. This indicates that the dark component makes the configurations more compact, consistent with the behavior already found in dark-matter admixed white-dwarf models \citep{Carvalho2025,Sahoo2026-ih}. For comparison purposes, observational data from the Montreal White Dwarf Database (MWDD; \citep{MWDD}) are displayed as black dots, with radii calculated directly from the cataloged mass and surface gravity. To ensure a sample of confirmed WDs, non-degenerate and unconfirmed objects (such as pre-degenerates, subdwarfs, and candidates) were excluded, retaining only confirmed spectral types (DA, DB, DC, DQ, DZ, and DO). Sources with missing required physical parameters were also removed, resulting in a sample of 133,704 objects. It is worth noting that besides error bars some small radii WDs could be explained by the inclusion of DM effects.  

A notable result is that the maximum-mass configurations remain in the white-dwarf regime but are systematically shifted toward higher compactness as the dark-matter fraction increases. The additional gravitational contribution of the dark component modifies the hydrostatic equilibrium, displacing the mass-radius sequences toward lower masses and smaller radii. As a consequence, the stellar configurations become progressively more compact, in agreement with the qualitative behavior shown in the figures.

\begin{figure}
    \centering
    \includegraphics[width=1.0\linewidth]{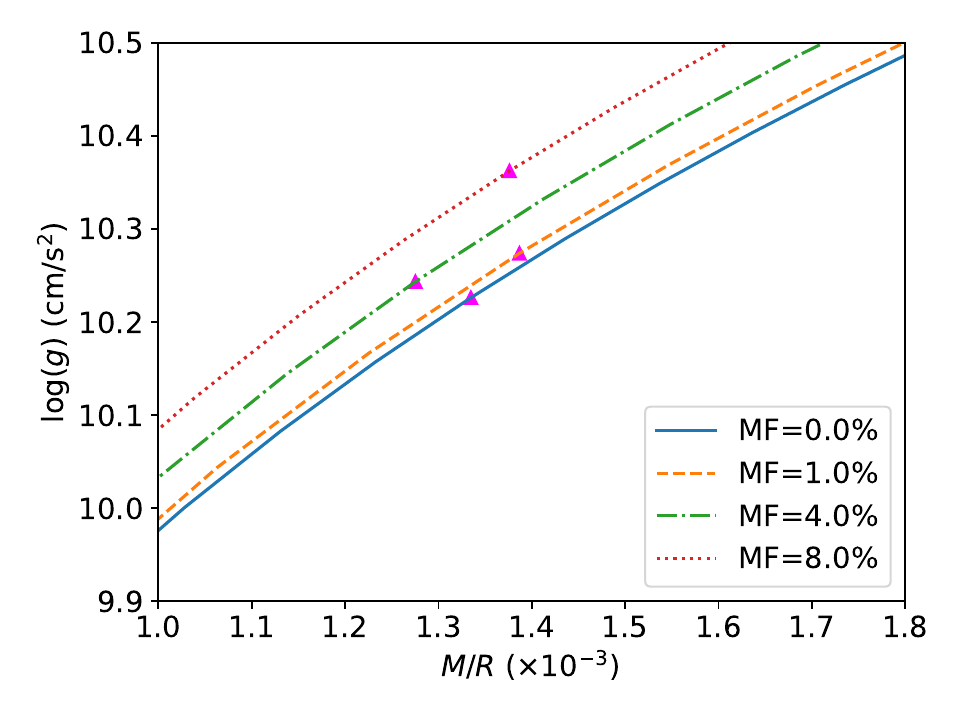}
    \caption{Surface gravity as a function of the dimensionless compactness for WDs for different quantities of DM. The full pink triangles mark the maximum mass points.}
    \label{gXcompactness}
\end{figure}

The surface gravity as a function of compactness is plotted in Fig. \ref{gXcompactness} for four different dark-matter fractions. In the figure, the pink triangles mark the maximum-mass configurations. In all cases, the surface gravity grows monotonically with compactness. For a fixed value of the compactness, we find that equilibrium configurations with larger dark-matter fractions have larger surface gravity. The figure demonstrates that increasing the dark-matter mass fraction inside WDs leads, in general, to higher surface gravities for the corresponding maximum-mass configurations.

This trend provides an observational bridge to spectroscopic measurements. In principle, for a fixed mass, WDs with high surface gravity could be interpreted as compact configurations with an internal dark component. This result is compatible with the discussion in \cite{Carvalho2025}, where it is emphasized that DM can induce high compactness and strong surface-gravity anomalies. In the present model, this effect is preserved even when the dark-matter content is parameterized by a fixed mass fraction.

\begin{figure}
    \centering
    \includegraphics[width=1.0\linewidth]{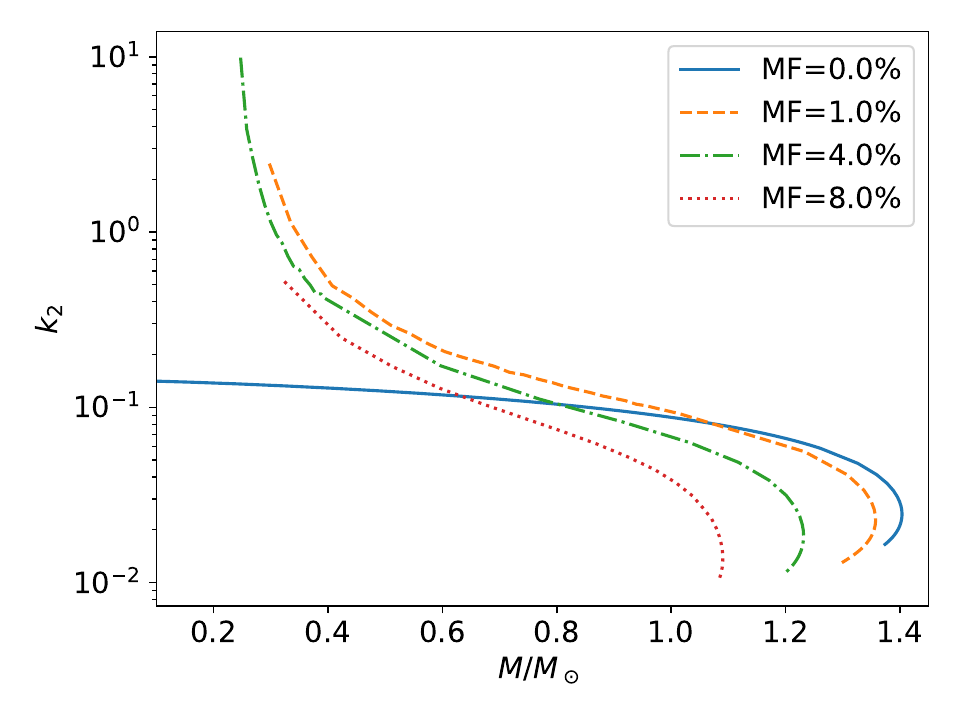}
    \caption{Love number-total mass relation for WDs with four different quantities of DM.}
    \label{massxk2}
\end{figure}

Figure \ref{massxk2} shows the profile of the Love number $k_2$ with stellar mass for different DM fractions. All curves exhibit a monotonic suppression of the Love number with increasing total mass. The effects of the DM fraction can be observed by analyzing the Love number. The inclusion of DM systematically reduces $k_2$ for a given mass and shifts the entire sequence toward lower masses. As a consequence, stars become less deformable under an external tidal field, which is reflected in the lower values of $k_2$. This behavior is expected because the DM contributes additional gravity while following a pressure distribution different from that of the white dwarf's baryonic matter. The result is a more compact configuration with a smaller quadrupole response. In physical terms, the mixing of DM in WDs reduces the star's ability to develop a large tidal quadrupole under an external field. This trend is consistent with the inverse correlation between compactness and $k_2$ found in tidal studies of compact objects \citep{Nunes2025,2008ApJ...677.1216H}.

\begin{table}[!ht]
\centering
\caption{Mass intervals for the primary ($M_1$) and secondary ($M_2$) WDs corresponding to selected chirp masses. All masses are given in units of $M_\odot$.}
\label{tab:mass_intervals_chirp}
\begin{tabular}{ccc}
\hline\hline
$\mathcal{M}\,[M_\odot]$ &
$M_1\,[M_\odot]$ &
$M_2\,[M_\odot]$ \\
\hline
$0.3$ & $[0.3476,\;1.4000]$ & $[0.1102,\;0.3416]$ \\
$0.5$ & $[0.5757,\;1.4000]$ & $[0.2668,\;0.5730]$ \\
$0.7$ & $[0.8070,\;1.4000]$ & $[0.4871,\;0.8012]$ \\
$0.9$ & $[1.0351,\;1.4000]$ & $[0.7766,\;1.0326]$ \\
$1.1$ & $[1.2664,\;1.4000]$ & $[1.1428,\;1.2607]$ \\
\hline\hline
\end{tabular}
\end{table}

\begin{table*}[!ht]
\centering
\caption{Selected mass fractions with correspondent minimum masses for positive $k_2$ and maximum masses with related radii, surface gravity and dimensionless compactness.}
\label{tab:mass_parameters}
\begin{tabular}{cccccc}
\hline\hline
Mass Fraction (MF) &
$M_{\rm min}$ ($M_\odot$) &
$M_{\rm max}$ ($M_\odot$) & 
R (km) & 
log($g$) (cm/s$^2$) & 
Compactness ($10^{-3}$)\\
\hline
0\% & -     & 1.404 & 1018.1 & 10.25 & 1.379 \\
1\% & 0.297 & 1.357 & 978.9  & 10.27 & 1.387 \\
4\% & 0.247 & 1.232 & 965.7  & 10.24 & 1.275 \\
8\% & 0.324 & 1.090 & 792.3  & 10.36 & 1.376 \\
\hline\hline
\end{tabular}
\end{table*}

It is worth noting that the smallest-mass points in each $k_2$ curve represent the limit for positive $k_2$. In fact, solutions with smaller masses yield negative values of $k_2$, mainly because the DM content extends farther than the normal baryonic matter content. This leads to solutions with $y_R<-3$, consequently yielding a negative Love number. A negative Love number means that the induced quadrupole has the opposite sign to the applied tidal field, i.e., the quadrupole is oriented in the reverse direction relative to the tidal field. In principle, this does not necessarily imply an instability; however, this behavior deserves further investigation, which will not be carried out here.

\begin{figure}
    \centering
    \includegraphics[width=1.0\linewidth]{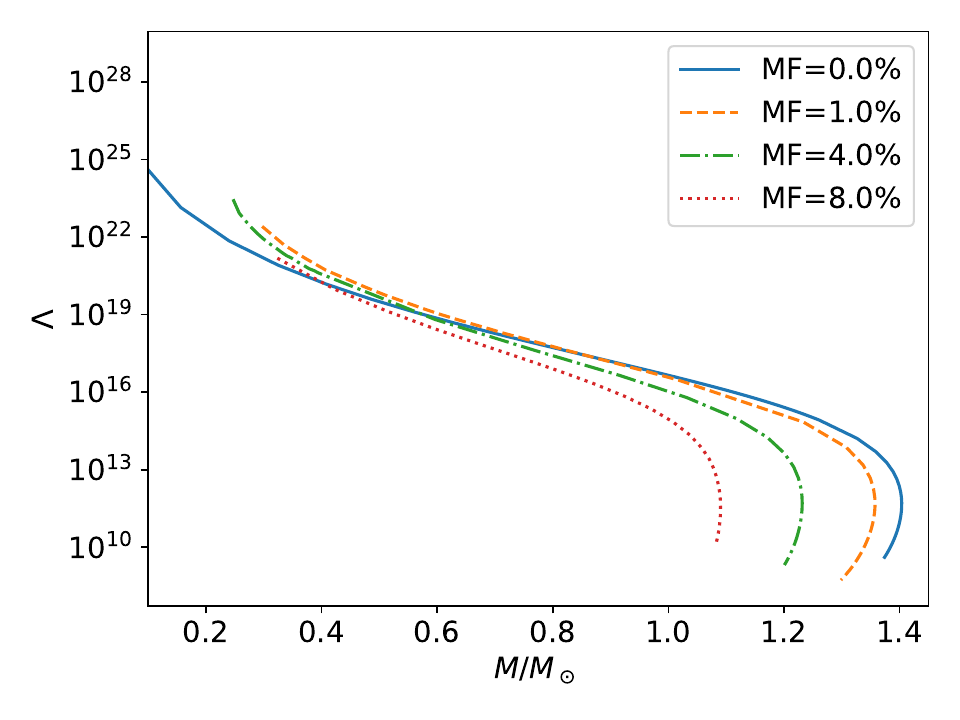}
    \caption{Dimensionless tidal deformability versus the total mass for some DM values.}
    \label{massxlambda}
\end{figure}

Figure \ref{massxlambda} shows the tidal deformability profile as a function of mass for different DM fractions. Note that the values obtained for tidal deformability are higher than those reported for the Love number; this is because the dimensionless deformability is given by $\Lambda=(2/3)k_2C^{-5}$, so the factor $C^{-5}$, associated with the low compactness of WDs, significantly amplifies the values of $\Lambda$, even when $k_2$ remains on the order of unity or less. The figures show a decrease in $\Lambda$ as mass increases, and the presence of DM accentuates this trend. Configurations with higher DM fractions are systematically less deformable under tidal perturbations for the same stellar mass.

This is one of the main observational implications of the model. If WDs containing a non-negligible amount of DM exist in nature, they would produce smaller tidal signatures during binary evolution and weaker finite-size effects in gravitational-wave phasing. In the context of double white dwarf systems, this would alter the inference of stellar radii and internal composition from the inspiral signal. The suppression of the tidal deformability therefore constitutes a direct fingerprint of DM inside the star \cite{Wolz2020-mh}.

\begin{figure}
    \centering
    \includegraphics[width=1.0\linewidth]{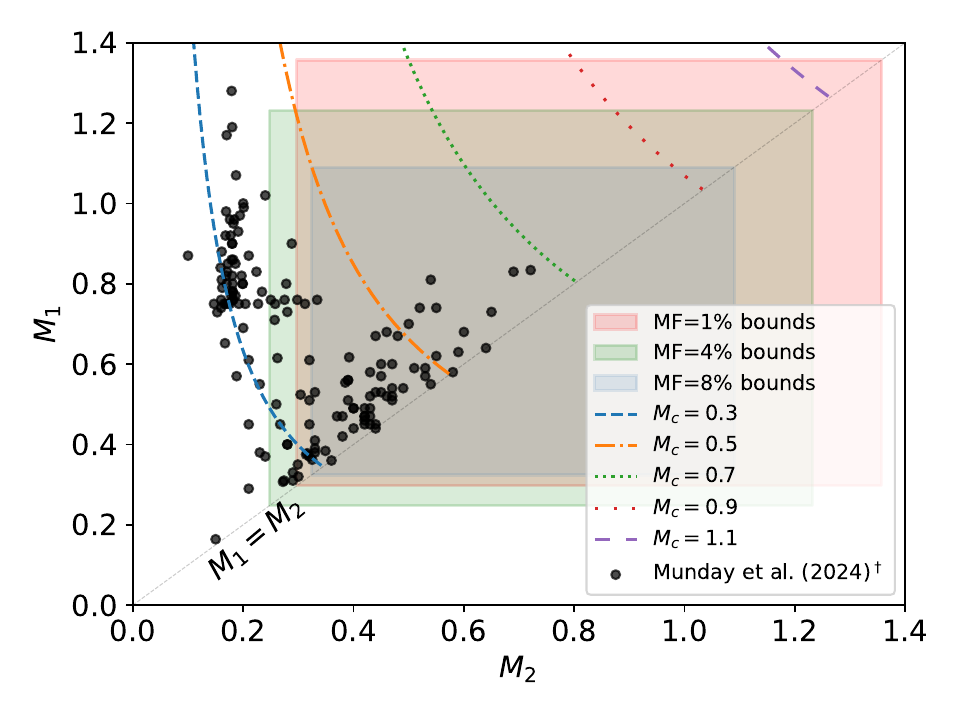}
    \caption{The $M_1-M_2$ plane for different chirp masses and DM quantities. The solid diagonal line denotes the equal-mass cases, $M_1=M_2$. The black dots represents observational data taken from \citep{Munday2024}. $^\dagger$The complete sample of close double white dwarf binaries can be found in the repository \citep{CloseDWDbinaries}.}
    \label{M2-M1}
\end{figure}

\begin{figure}
    \centering
    \includegraphics[width=1.0\linewidth]{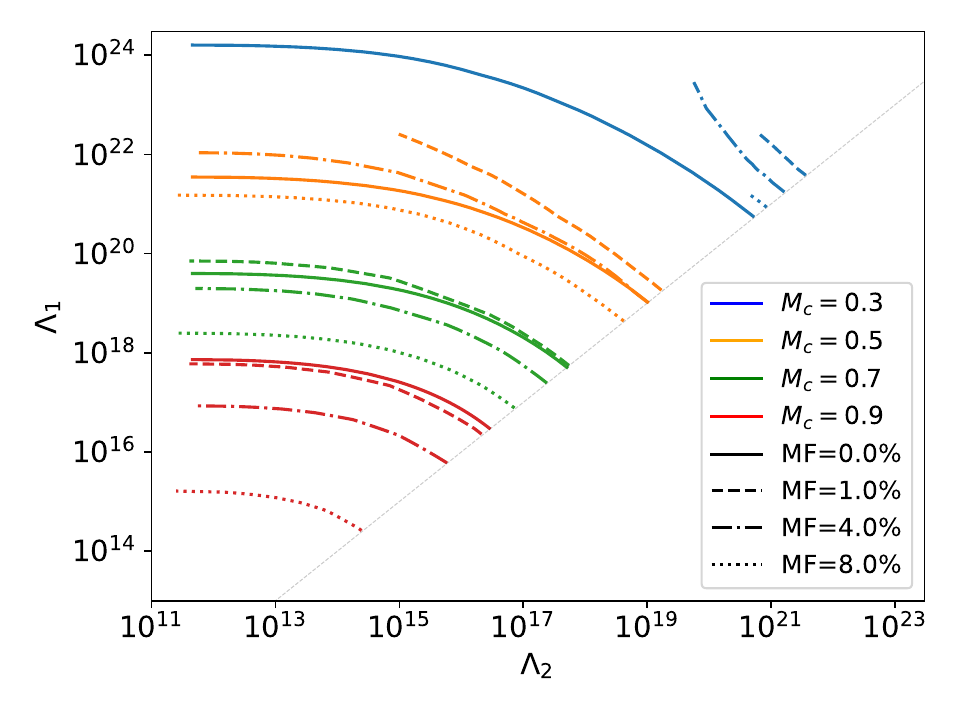}
    \caption{The $\Lambda_1-\Lambda_2$ plane for binary WDs with different chirp masses and DM fractions. The diagonal line marks the equal tidal deformability configurations case, $\Lambda_1=\Lambda_2$.}
    \label{L2-L1}
\end{figure}

In the low-order post-Newtonian approximation, the chirp mass provides a compact and physically meaningful way to describe a binary system composed of two compact objects with masses $M_1$ and $M_2$. Rather than treating the two masses independently, the chirp mass is introduced because it governs the leading-order evolution of the orbital frequency and therefore sets the dominant scale of the gravitational-wave signal during the inspiral phase. For this reason, it is the key mass parameter most directly constrained by the observed waveform \citep{Abbott2016GW150914,Abbott2016PropertiesGW150914,Abbott2017GW170817,Abbott2017MultiMessengerGW170817,Abbott2018TidalGW170817,Abbott2019GWTC1,Abbott2021GWTC2,Abbott2023GWTC3}.

In practical applications, the component masses are not completely arbitrary, since the labels are conventionally assigned such that $M_1\geq M_2$. This ordering does not change the physics, but it helps avoid ambiguity when constructing tidal deformability diagrams. Under this convention, the chirp mass and mass ratio together provide a convenient and compact parametrization of the binary, capturing the essential dynamical information while simplifying the analysis. The $M_1$--$M_2$ diagram presented in Fig. \ref{M2-M1} is obtained by solving the chirp-mass equation for a fixed value for $M_c$,
\begin{equation}
    M_c= \frac{(M_1M_2)^{3/5}}{(M_1+M_2)^{1/5}}.
\end{equation}

The diagonal curve in Fig. \ref{M2-M1} represents $M_1=M_2$ which separates the physically relevant region $M_1>M_2$. The black dots in the figure highlights the observational data taken from the close double WDs binaries catalog, which is a maintained list, with reference and details, of compact double WDs (CDWD) by \cite{Munday2024}. The complete and updated list of CDWDs can be found in the repository \cite{CloseDWDbinaries}. In Fig. \ref{M2-M1}, we exclude binaries with absent data for $M_1$ or $M_2$, and also data with only lower limits on $M_2$. The final sample contains 152 selected binaries. The shaded regions for each  mass fraction represent the minimum and maximum masses of curves presented in Fig. \ref{massxk2}. Table \ref{tab:mass_intervals_chirp} lists the mass intervals for $M_1$ and $M_2$ used to generate the $M_2$-$M_1$ diagram presented in Fig. \ref{M2-M1}. Table \ref{tab:mass_parameters} lists the minimum mass for positive $k_2$ and the maximum masses with correspondent radii, surface gravity and compactness. 

The $\Lambda_1$--$\Lambda_2$ diagram presented in Fig. \ref{L2-L1} shows a systematic suppression of the tidal response in both components of the binary. The curves associated with larger values of MF lie below the ordinary-white-dwarf case, indicating that the presence of DM makes both stars less deformable. This behavior is expected since the DM component modifies the stellar structure, changing the density distribution and reducing the tidal response of the system.

Since tidal effects in a binary depend on the deformabilities of both components, the shifts observed in the $\Lambda_1$--$\Lambda_2$ relation may affect the gravitational-wave phase evolution. In particular, the tidal contribution is determined by a mass-weighted combination of $\Lambda_1$ and $\Lambda_2$, and therefore a simultaneous suppression of both parameters can lead to a smaller tidal correction relative to an ordinary double-white-dwarf binary with the same component masses. Such a modification may also introduce a bias in the inference of the binary parameters if the waveform is analyzed assuming ordinary WDs.

For the larger chirp-mass case, $M_c=0.9~M_\odot$, the dark-matter curves can differ from the ordinary-white-dwarf case by up to approximately three orders of magnitude. For smaller chirp masses, the deviations are less pronounced, but the tidal deformabilities still change by at least one order of magnitude for some configurations. Note also that, for low binary masses ($M_c=0.3$), the $\Lambda_1-\Lambda_2$ curve spans different allowed ranges in the no-DM case compared with the DM scenarios. These differences indicate that dark-matter admixed WDs may leave a relevant imprint on the tidal sector of the gravitational-wave signal.

The possible detectability of these effects by LISA depends not only on the relative difference in $\Lambda_1$ and $\Lambda_2$, but also on the binary frequency, signal-to-noise ratio, distance, observation time, and degeneracies with the component masses, spins, tidal synchronization, and mass-transfer effects. Tidal signatures are expected to be more favorable for high-frequency double-white-dwarf binaries, where the accumulated phase shift is larger and LISA can more accurately measure the frequency evolution. Studies of tidal coupling in double WDs show that tidal corrections can be measurable for sufficiently loud systems at frequencies in the mHz regime, while neglecting them may bias the estimation of source parameters \citep{Fiacco2024-ie}.

Therefore, the large deviations found here, especially for the high-chirp-mass configurations, suggest that DM effects could in principle be distinguishable from the ordinary-white-dwarf case in a favorable LISA observation. However, a definitive statement about detectability requires a dedicated parameter-estimation analysis using waveform models that consistently include the dark-matter-modified tidal deformabilities, as well as other relevant finite-size and binary-evolution effects.

Taken together, the binary diagrams reinforce the message of the isolated-star analysis: DM reduces tidal deformability and increases compactness. This makes the stars less susceptible to tidal distortion but harder to distinguish from standard WDs without high-precision measurements.

\section{Conclusions}\label{sec4}

We studied the tidal properties of WDs admixed with fermionic DM using a two-fluid relativistic formalism and a fixed dark-matter mass fraction, ${\rm MF}=M_{\rm DM}/M_{\rm WD}$. By combining the dark-matter framework of \cite{Carvalho2025} with the tidal-deformability formalism, we obtained a consistent description of the structural and tidal properties of the stars. The resulting sequences show that DM systematically decreases the radius, increases compactness and surface gravity, and suppresses both the Love number and tidal deformability.

The binary diagnostics further indicate that DM shifts the $(\Lambda_1,\Lambda_2)$ relations toward more compact and less deformable configurations. These effects could have consequences for the interpretation of white-dwarf binaries in the context of gravitational-wave astrophysics and compact-star evolution. In particular, the model suggests that precision measurements of tidal signatures and surface gravity may provide indirect constraints on DM inside WDs.

Future work may extend this study by including more realistic equations of state for white dwarf matter description, self-interacting fermionic DM or even a different DM particle nature. It would also be interesting to combine the present static treatment with binary evolution and waveform modeling, especially in the context of upcoming low-frequency gravitational-wave missions.

\section*{Acknowledgments}

GAC would like to thank the financial support of CNPq (Conselho Nacional de Desenvolvimento Cient\'ifico e Tecnol\'ogico) under process \#314121/2023-4, Funda\c{c}\~ao Arauc\'aria under NAPI ``Fen\^omenos extremos no Universo'', and Funda\c{c}\~ao de Amparo a Pesquisa e Inova\c{c}\~ao do Esp\'irito Santo (FAPES) process 12/2024. JDVA would like to thank Universidad Privada del Norte and Universidad Nacional Mayor de San Marcos for the grant - RR No.$\,005753$-$2021$-R$/$UNMSM under the project number B$21131781$.  JGC is grateful for the support of FAPES (1020/2022, 1081/2022, 976/2022, 332/2023), CNPq (306018/2025-0), and FAPESP (grant No. 2021/01089-1).

%\appendix

%\section{Appendix title 1}

%\section{Appendix title 2}

\bibliographystyle{elsarticle-harv} 
\bibliography{example}



%% For example:

%% \bibitem[Aladro et al.(2015)]{Aladro15} Aladro, R., Martín, S., Riquelme, D., et al. 2015, \aas, 579, A101

%%\end{thebibliography}

\end{document}